\documentclass{caosp307}

\usepackage{graphicx}

\usepackage{natbib}
\articleNo{123}
\pubyear{2005}
\volume{35}
\volnumber{3}
\firstpage{1}
\received{May 1, 2005}
\accepted{August 28, 2005}

\def\BibTeX{{\rm B\kern-.05em{\sc i\kern-.025em b}\kern-.08em
             T\kern-.1667em\lower.7ex\hbox{E}\kern-.125emX}}

\begin{document}

%
\hauthor{M.\, Yu.\, Skulskyy}

\title{Formation of magnetized spatial structures in the Beta Lyrae system }
\subtitle{ I. Observation as a research background of this phenomenon}


%
\author{
M.\, Yu.\, Skulskyy
       }

%
\institute{
Lviv Polytechnic National University, Department of Physics, 79013, Lviv, Ukraine, \email{mysky@polynet.lviv.ua}
          }

\date{March 8, 2003}

\maketitle

\begin{abstract}
The discovery of the donor magnetic field has repeatedly posed the task of a thorough study of the phenomenon, which is based on the concept of the influence of the magnetic field on the processes of the formation of gaseous structures and mass transfer in the Beta Lyrae system. This article provides an overview, analysis, and synthesis of the results of a variety of long-term observations as a necessary basis for further clarification of issues aimed primarily at the study of magnetized gaseous structures. As a part of such a study, it was found that the structure of the gaseous flows between the donor and the gainer varies in some way depending on the phases of the orbital period; and, accordingly, that the donor magnetic field influences the formation of these moving magnetized structures. The analysis of the masses of both components for use in further scientific works suggests that the following values are optimal: 2.9\,$M_{\odot}$ for the donor and 13\,$M_{\odot}$ for the gainer. The study of satellite lines as a certain phenomenon leads to the fact that the accretion disk surrounding the gainer consists of two parts: the external satellite disk and the internal massive opaque disk. From the analysis of all observations and studies of the magnetic field, observations on the 6-m telescope can be considered the most reliable. They have formed the spatial configuration of the donor magnetic field, which is important for studying and understanding the features of the mass transfer in this interacting system. Further evidence regarding the picture of the magnetized accretion structures as the special phenomenon will be presented in the following articles.
\end{abstract}

%
\section{Introduction}
\label{intr}
Our spectral and spectrophotometric observations of the known interacting binary Beta Lyrae were conducted in 1965-2010 on different telescopes, in particular, on 6-m and 2.6-m telescopes at the Special and Crimean Astrophysical Observatories with using CCD detectors and polarization analyzers. The most important our results are the first definition of the masses of both components directly from the complex spectra of the binary system \citep{Skulskij1975a, Skulskij1992}, simultaneously, with the discovery and research of the donor’s magnetic field \citep{Skulskij1985a, Skulsky1993}, as well as the study of the structure of the accretion disk surrounding the gainer \citep{Skulskij1993c, Skulskij1993a, Skulskij1993b}. The results of the last two directions of our observations and researches are outlined in various editions forming certain ideas about the basic physical properties of the Beta Lyrae system, which is in the phase of the rapid mass transfer. But, they are practically not used in another, methodologically dissimilar, studies in interpreting its nature. This applies, e.g., of the articles \cite{Mennickent2013} and \cite{Mourard2018}, where the progress in the modelling of light curves of the Beta Lyrae system was made, in particular, the significant contribution of the radiation of the accretion disk to the light curves have been established.
 
In this sense, the identification of two hot regions (or spots) on the accretion disk in \cite{Mennickent2013} was of particular interest. These regions are observed in the phases of 0.80\,P and 0.40\,P and their temperatures are 10\% and 20\% higher than the average on the disk. The authors supposed that these hotter regions illuminated by the donor may be formed by the collision of gas flows with the disk during mass transfer between the components of this system (but without mechanism suggested for both hot regions). At the same time, the results of this light curve modeling can be interpreted in the light of studies of the accretion disk properties at the presence of the donor magnetic field. Indeed, the hotter region of the accretion disk at the phase 0.40\,P of the first quadrature is naturally explained to the deflection by the Coriolis force of the main gas flow that is directed from a donor through a Lagrange point to the gainer's Roche cavity and with a further collision of this flow with the accretion disk. This hotter region and its location are well known. In particular, it was clearly detected by \cite{Burnashev1991B} absolute spectrophotometry of the Beta Lyrae system and followed from various spectropolarimetric observations. However, such a classical hydrodynamic picture cannot be suitable for explaining of the heated part of the accretion disk in the orbital phases near 0.80P of the second quadrature. Here, this should be based on another understanding of the sources of this disk heating and the structure of the accretion flows between the components of this interacting binary system, given the presence and influence of a donor magnetic field. Such our understanding was formed mainly in a number of original observations and investigations in 1980\,--\,1995. 

In this situation, there was a need for a consistent generalized statement of certain aspects of our spectral studies based on the concept of the reality of magnetized gaseous structures that reflect the specific configuration of the donor magnetic field. The photometry and absolute spectrophotometry data, orbital variability of donor magnetic field curves, radial velocities and intensities of complex lines in different spectral regions, etc. were analyzed. As a first result, the preliminary data to a pattern of the accretion flows was briefly represented by \cite{Skulsky2015, Skulsky2018}. However, due to a large increase in data in the study of diverse observations, this summarized research had to be divided into three related articles. The purpose of this article is to give a brief description of both our spectral observations and other diverse studies as a prerequisite for creating a general mass transfer picture in the presence of a donor magnetic field.

\section{Introductory description of properties and a schematic picture of the Beta Lyrae system}
\begin{figure}[!t]
	\centerline{\includegraphics[width=0.75\textwidth,clip=]{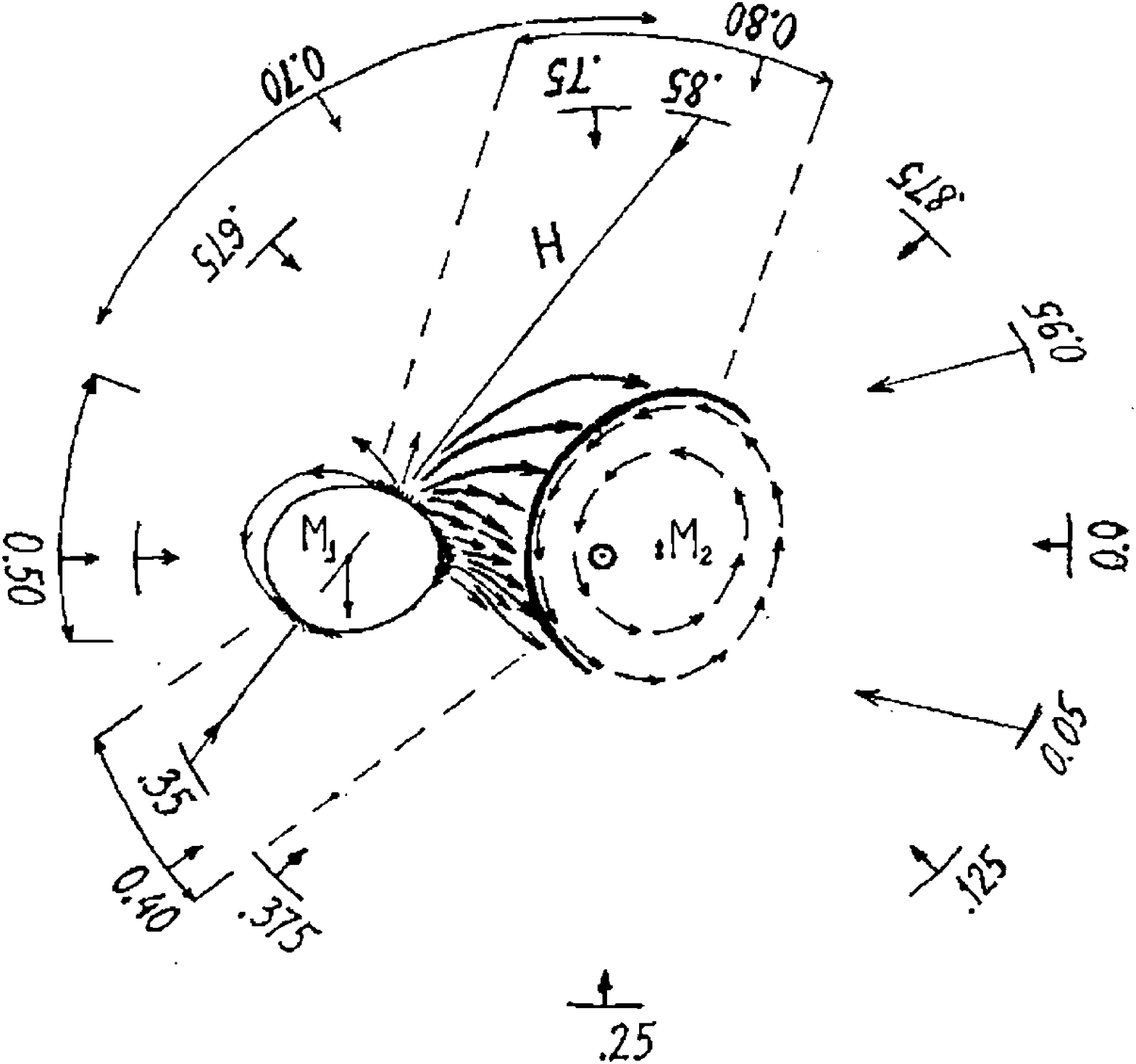}}
	\caption{ Schematic model of the Beta Lyrae system (top view on the orbital plane). It contains the bright donor (reaches its Roshe lobe) of mass $2.9 M\odot$ and the gainer of mass $13 M\odot$ masked by an opaque accretion disk. 
	The distance between centers of two components on the real scale is $58 R\odot$ (small circle shows the gravity center of the binary system).
	More important phases (as fractions of the orbital period), the directions of the orbital motion of two components, and the dipole axis of the donor effective magnetic field (denoted by "H") shown by straight arrows. The fan-shaped gaseous accretion flows, which represented by curved arrows, form a hot arc (bold line) on the outer rim of the accretion disk facing the donor. The dimensions of the outer part of the accretion disk (so-called satellite-disk) represented by two rows of round arrows in the direction of its rotation.
	This rotation in the direction towards and from the observer shown respectively in phases 0.05\,P and 0.95\,P (phase 0.0\,P corresponds to the center of the primary eclipse). Two hotter regions on the disk, proposed at the modeling of light curves by \cite{Mennickent2013}, are marked in the phases of 0.40\,P and 0.80\,P by dashed limits directed at the centers of both components.}
	\label{fig_scheme}
\end{figure}

Preliminary data from this large study suggest that a further generalized description and investigation of the physical properties of the Beta Lyrae system will be better comprehensible when before will submit a brief description of the verified facts and some illustration. 

According to the absolute spectrophotometry by \cite{Burnashev1978} and spectral observations by \cite{Skulskij1992, Skulskij1993c}, one can assume (see Fig.\,\ref{fig_scheme}) that interacting system Beta Lyrae in the range of  $\lambda \lambda~ 3400-7300 ~\AA$ contains the bright donor of type B8III of mass 2.9\,$M_\odot$, the massive gainer of mass 13\,$ M_\odot$ masked by an opaque thick accretion disk with a pseudophotosphere of A5III type, and the hydrogen envelope with temperature close to $T_e = 20000 K$. The chemical composition of the atmosphere of the donor is peculiar and undergoing hydrogen burn up in the CNO cycle \citep{Skulskij1975b, Skulskii1986, Balachandran1986}. It is believed that the donor practically fills its Roshe cavity and is being in the final stage of mass loss and transfer, the rate of which is close to $2\cdot 10^{-5} M_{\odot}/yr$. 
As Umana (2002) noted, the radio emission from the binary system supports the presence of an extra source of continuum emission generated by stellar winds with a mass loss rate of 5\,$\cdot10^{-7} M_{\odot}/yr$ In this binary system, there are well-developed circumstellar structures, in particular, of the accretion disk surrounding the gainer. This disk is formed by at least two of its components \citep{Skulskij1992, Skulskij1993c}: an outer translucent so-called satellite disk, whose rotational effect is clearly apparent in its projection on the donor at the phases only before and after the donor eclipse, and an optically thick opaque massive disk (approximately of ($10^{-4}-10^{-3}$)\,$M_{\odot}$ by Mourand et al., 2018).
At measurements of hundreds of Zeeman spectrograms, a magnetic field was detected in the donor atmosphere, which quasi-sinusoidally changes over of orbital period phases \citep{Skulskij1982, Skulskij1985a}. 
A dipole magnetic field with an axis directed in the direction of orbital phases of (0.355-0.855)\,P has a maximum on the donor surface at the 0.855\,P phase reflecting the location of this magnetic pole facing the gainer. This meant that the ionized gas, channeled by the donor magnetic field, can moves in the direction of its dipole axis from the donor surface and deflects along the magnetic field lines toward the accretion disk \citep{Skulsky2015, Skulsky2018}. The correlation between the phase variability in the absolute flux in H emission line and such in the effective magnetic field of the donor discovered \cite{Burnashev1991B}. This was the first substantial fact to support the concept that the magnetic field has a significant effect on the spatial formation of gaseous structures during mass transfer in the Beta Lyrae system. This comprehension confirmed by the results of further observations. It is hoped, a better understanding of the basic research in this article will be facilitated by the presentation in Figure 1 of a schematic model of this binary system.


\section{Results of the most significant observations and investigations of the Beta Lyrae system}
This article is not intended to be a detailed review of all Beta Lyrae publications, a full description can be found in \cite{Sahade1980}. Recent observations, especially regarding the topic of modern modeling of light curves, were reported, for example, in \cite{Mourard2018}. The following sections will give a more detailed overview of the important observations, the results of their research and the interpretation that are related to the coverage of the declared topic. To illustrate some observed facts, we have sometimes to use original drawings from the articles cited. 
      
\subsection{Photometry of the Beta Lyrae system and simulations of its light curves}
Numerous photometric publications about the Beta Lyrae system affirm a constant interest in its physical nature. Starting from the observations of \cite{Goodricke1785}, the orbital period increased by 0.05\,d and now is reaching of 12.94\,d. This clearly indicates the phase of rapid mass transfer between both components of this binary system. It characterized by a high rate of about 2\,$\cdot10^{-5} M_{\odot}/yr$, whereas the age of this massive interacting system of near 2.5\,$\cdot10^{-7} M_{\odot}/yr$ years is relatively young. As a result, well-developed gaseous structures are observed in this binary system (a certain presentation of its is given in Fig.\,\ref{fig_scheme}). That is why the main focus in modern modeling of light curves of the Beta Lyrae system is paid to the radiation of the circumstellar structures, especially the accretion disk, and their physical properties. 

\begin{figure}
	\centerline{\includegraphics[width=0.75\textwidth,clip=]{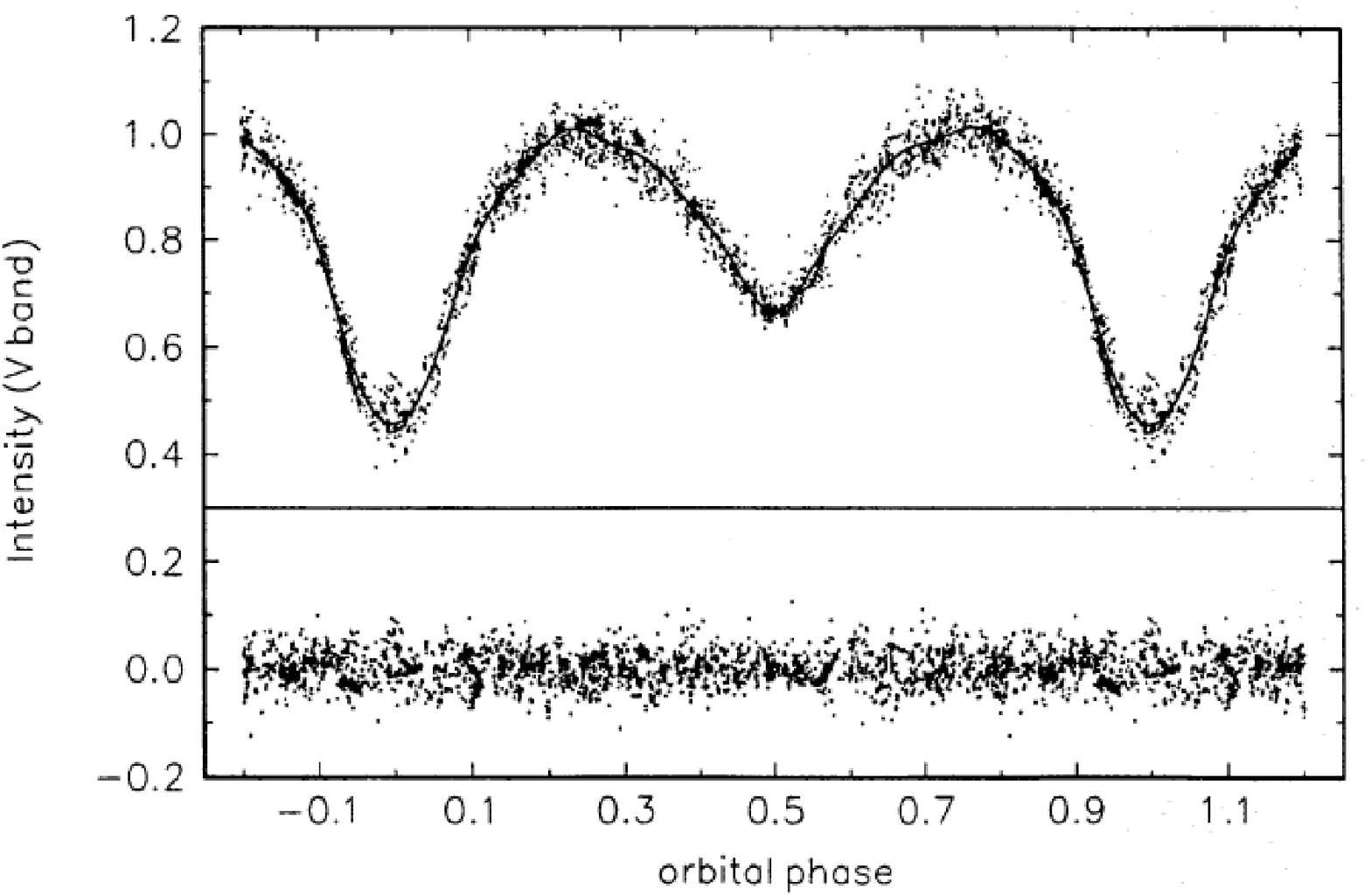}}
	\caption{V-band light curve ($\lambda~5500 ~\AA$) \citep{Harmanec1996}}
	\label{b_lyrae_V}
\end{figure}

The well-known "light curve of type Beta Lyrae" in modern observations, presented in Fig.\,\ref{b_lyrae_V}, was published by \cite{Harmanec1996}. Their data set is often used in light curve simulations based on the idea of an optically thick accretion disk surrounding the massive gainer. 

According to a number of light curve simulations, including recent studies \cite{Mennickent2013} and \cite{Mourard2018}, it can be considered that this opaque disk starts from the gainer with its radius in $6 R_{\odot}$ at the semi thickness of the disk that is about 6.5\,$R_{\odot}$, the radius of the outer rim reaches 30\,$R_{\odot}$, and the binary orbital inclination is in ranges from $81\degr$ to $93.5\degr$. The authors indicate a significant contribution of the radiation of this disk to the total light curve from the side of the disk facing the donor. Although the idea of heating this part of the disk, including an interaction with the falling gas flows, is considered in many models, there are new aspects. The radial temperature profile of the disk varies from 30000\,K to 8000\,K, where the central regions of the disk are heated by the embedded gainer. On outer edges this disk their surface temperatures is essentially different, particularly from the side of the disk facing the donor where there is the wide asymmetric hotter region with the temperature that can reach 10000\,K. The light-curve model of \cite{Mennickent2013} has shown two hotter regions in the phases 0.80P and 0.40P with covering respectively 30\% and 10\% of the accretion disk edge. Their temperatures are higher on 10\% and 20\% from the mean value on the disk edge. The wide bright region in the phase 0.80\,P concerns of the second quadrature, in which the accretion disk contributes 22\% to the total V-band light curve (that is matched with the Beta Lyrae absolute spectrophotometry of \cite{Burnashev1978}). 

\begin{figure}[t]
	\centerline{\includegraphics[width=0.75\textwidth,clip=]{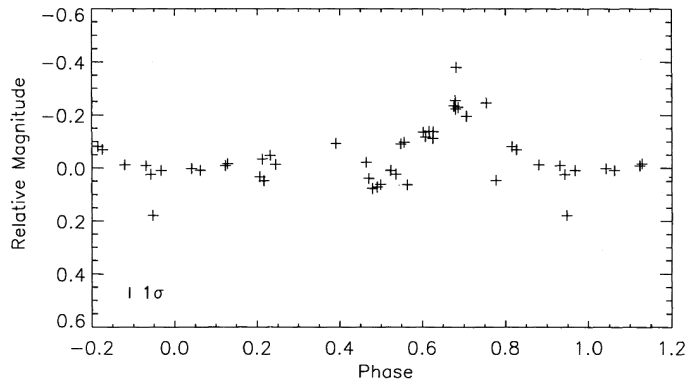}}
	\caption{Ultraviolet light curve ($\lambda~1085 ~\AA$) \citep{Kondo1994}}
	\label{b_lyrae_U}
\end{figure}

Examining a question of the different geometric shapes of the accretion disk, in \cite{Mourard2018} considerable progress in the simulations of the light curves from the far ultraviolet to the infrared region has been made. However, there are some difficulties in harmonizing the results of the modeling of light curves relating to the bands where the crossover is observed at depths of minima (see, for example, the band curves at 1250\,$\AA$ and 1365\,$\AA$ in \cite{Kondo1994}). This is especially true of the spectral region at wavelengths shorter than the Lyman limit, where, according to \cite{Kondo1994}, the band curves at 965\,$\AA$ and 1085\,$\AA$ show that the Beta Lyrae system here looks not at all as an eclipsing binary. These Voyager satellite observations have shown that light curves outside the Lyman limit are deformed in such a way that they lose their characteristic features that are typical for the visible spectrum shown in Fig.\,\ref{b_lyrae_V}. The ultraviolet light curve at 1085\,$\AA$ (see Fig\,\ref{b_lyrae_U}, which borrowed from \cite{Kondo1994}) becomes almost flat with a clear maximum of the radiation in the second quadrature. Such a light curve at 965\,$\AA$ in the same second quadrature looks as quasi-sinusoidal with the observed outburst lasting several days. 
Hence, that in the (0.6-0.8)\,P phases of the second quadrature (see Fig.\,\ref{fig_scheme}), to the left of the direction (0.5-1.0)\,P, the dominant region of shock collision of the hot turbulent plasma with the accretion disk is observed, reflecting the spatial configuration of the donor magnetic field \citep{Skulskij1985a}.

Understanding the nature of the Beta Lyrae system based on the light curve solutions requires taking into account all available data. An important factor is the existence of the soft X-ray emission (below 10\,keV), which does not change with the phase of the orbital period and is not obscured by both components. According to \cite{Ignace2008}, this soft X-ray emission must radiate over the orbital plane, probably due to the 
Thomson scattering of X-rays in the stellar wind of the hot gainer or this may be due to shock collision gaseous flows with the accretion disk. This approach is supported by the observation of the radio-nebula, surrounding the Beta Lyrae system up to 40\,AU, which indicates the substantial non-conservative mass-loss during the evolution of this interacting system \citep{Umana2000}.

\subsection{The 283 d photometric periodicity and the coherent system of secondary periods}
The question of the reality of secondary non-orbital periods in the Beta Lyrae system and the nature of the corresponding periodic processes are discussed over a hundred years. The ranges of changes of periodicities and resonances of physical parameters are observed from tens of seconds to tens of years. Particular attention deserves non-orbital of 283\,d photometric periodicities established quite reliably in two consecutive articles. His first definition of ($283.39\pm0.26$) d was obtained by \cite{vanHamme1995} based on the deviations from the averaged light curves for the past 150\,years of the Beta Lyrae observations that are centered about 1920. In a year, this period was determined by \cite{Harmanec1996} with even greater accuracy ($282.405\pm0.070$)\,d, based on the averaged V-curve. It consists of 2582 V-band magnitudes obtained in a 36 yr period inclusive to 1989 and is presented in Fig.\,\ref{b_lyrae_V}. At the same time, at investigating the variability of the emission lines \cite{Harmanec1996} the secondary period of 4.478 d was found. It turned out that this short period, the orbital period, and the period 282.425 d are related by a simple equation. In addition, \cite{Peel1997} shows that in periodograms based on observations between 1898 and 1916, the peaks of periods of 283\,d and 340\,d are clearly visible, with a significantly stronger peak of 340\,d. \cite{Burnashev1980} based on the narrowband absolute spectrophotometry of Beta Lyrae in 1978-1979 derived quasiperiodic variations of the radiation flux in $H_{\alpha}$-line emission being 1/7 of the orbital period. 

The growth of the orbital period owing to the rapid mass transfer from donor to gainer logically results in the idea of coordinated variability in time of all known secondary periods. This topic was considered by \cite{Skulsky2000} in view of all sides of a question on the donor as a magnetic rotator. 
There was presented a mutually agreed system of periodicities that included already known periods with new periods of 10.93\,d, 13.23\,d, and 564.85\,d. In particular, the difference of 0.965\,d between the definitions of a long period in the articles of \cite{vanHamme1995} and \cite{Harmanec1996} was simply explained by two different sets of observational data, the effective centers of which were shifted in time by about 60 years. It turned out that studies of all secondary periods should be carried out in conjunction with the research of the magnetic field and the spatial structure of the accretion disk \citep{Skulsky2000, Skulsky2007}. And this created the prerequisites for the detection of a coherent system of secondary periods and resonances, which changes synchronously with the growth of the orbital period. It has been established that the long period of 283\,d can be a proof of the synchronization between the precession rotation of the outer part of the accretion disk (i.e., of satellite-disk, see Fig.\,\ref{fig_scheme}) and the tidal wave on the surface of the donor. The 283\,d, 340\,d, and 565\,d periods may reflect variable substructures of outer layers of the accretion disk, whose dynamic parameters are related to each other as integers. 

\subsection{Phenomenon of the satellite lines and physical properties of the accretion disk}
The original feature of the absorption spectrum of the Beta Lyrae system, unusual for interacting systems of the W Ser type, is the so-called satellites of absorption lines. They appear near some absorption lines in the binary system spectrum in the phases of 0.1-0.9\,P of the main eclipse, disappearing within 0.02\,P in the center donor eclipse. The system of satellite lines changes a sign of their radial velocities to the opposite at the passage of the main eclipse. It is believed that the satellite lines are formed in the gas substance of the outer edges of the accretion disk when it is passing in front of a bright donor. In our long investigations of the Beta Lyrae spectrum and based on well-known studies, starting with \citep{Struve1941} we have proceeded from the interpretation of satellite lines based on the idea of the rotation effect of the outer layers of the accretion disk. Finally, there was a need for more purposeful research of the satellite lines system as a key to understanding the structure of the accretion disk. Therefore, in 1985-1992, such research became a priority since observations were carried out with high spectral resolution and signal-to-noise ratio on a 2.6-m CrAO telescope at using a CCD detector. The phenomenon of the origin, development, and disappearance of the satellite lines HeI, H, MgII and SiII in the phases of the main eclipse was studied on the basis of measuring their radial velocities and equivalent widths. Also, in these years, one conducted a comprehensive study of the lines of the donor and gainer and the emission in the SiII $\lambda \lambda$ 6347, 6371 doublet. The main results of our first study, published in \cite{Skulskij1992}, are as follows.

The movement of satellite lines in the Beta Lyrae spectrum indicates the spatial stratification of their formation in the accretion disk. The radial velocity curves of satellite lines belonging to different chemical elements differ by their amplitude and other characteristics due to the difference between excitation and ionization potentials. These curves are asymmetric relative to the main eclipse median. The helium satellite lines have the greatest amplitude of change (from +240 km/s in the phase 0.96\,P before the eclipse center to -290 km/s in the 0.02\,P phase after the eclipse center). The satellite lines of silicon show an amplitude of change, correspondingly from +215 to -280 km/s, while the magnesium lines show even smaller changes, respectively, from +205 to -250 km/s. Taking into account the velocity of the center of mass of the binary system of -18 km/s one can lead to the conclusion that the negative velocities in the disk after the center of the eclipse are slightly higher than their positive velocities before the eclipse center. This indicates the phase asymmetry of the dynamic characteristics of the satellite lines: the maxima of the radial velocity curves there are within phases of (0.035-0.045)\,P phases before the eclipse center 
(i.e., in the phase of 0.96\,P) and within phases of (0.01-0.02\,P) after this center. In the closest phases to the eclipse center (and to the center of the disk) one sees the maximum of the radial velocity curve for the helium satellites that indicates the higher temperatures of the medium their formation. 
The phase dependence of the equivalent width of the satellite lines indicates the limits of their occurrence and
disappearance in the phases up to (0.89\,P and 0.98\,P) and after (0.01\,P and 0.11\,P) of the eclipse center, as well as the respective phases of their maximum intensity at (0.95\,P and 0.05\,P). The satellite lines disappear and reappear asymmetrically relative to the median of the eclipse and show a complete absence in the spectrum in the range greater than 0.02\,P, indicating the boundaries of the internal asymmetric opaque disk.

\begin{figure}
	\centerline{\includegraphics[width=0.75\textwidth,clip=]{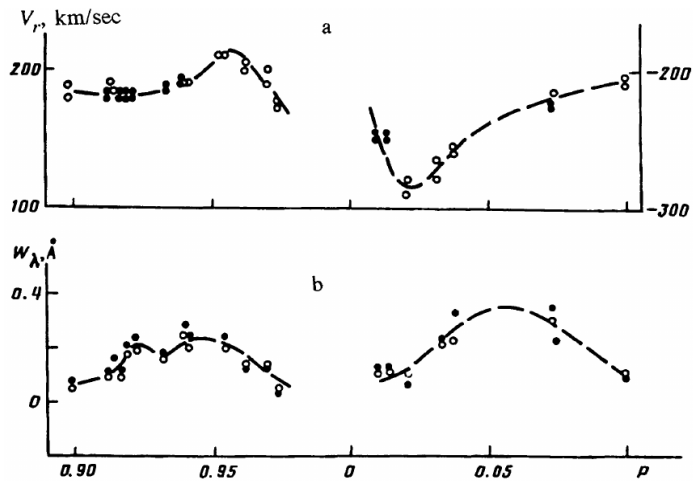}}
	\caption{a) Radial velocity curves of silicon line satellites, according to observations in 1991 (light circles) and 1992 (dark circles) ($V_r$ scales in phases before and after eclipse center are vertically offset. b) $W_\lambda = f(P)$ curves of line satellites of SiII $\lambda$ 6347 (filled circles) and SiII $\lambda$ 6371 (open circles) of equivalent widths enhanced at phase 0.92\,P. \citep{Skulskij1993a} }
	\label{b_lyrae_SiII}
\end{figure}

The revealed patterns of the phenomenon of satellite lines are shown in Fig.\, \ref{b_lyrae_SiII} and schematically reflected in Fig.\,\ref{fig_scheme} in the model of a satellite-disk, which has an elongated oval shape in the direction of the gainer's movement. The ratio of the axes of this disk is about 4/3. The greater half of the long axis shows some stratification for the disk layers with different chemical elements in the range of about 0.1 of the distance between the centers of the two components. On the backside of the gainer, where the radius of the disk is smaller, the rotating speed of its outer edges is higher. This is probably due to the action of the main gas stream that presses here the outer part disk to the surface of the gainer. Stratification of the conditions for the excitation of different satellite lines shows the increase in temperature and the acceleration of the rotation of disk layers to the center of the gainer. However, the speed of their rotation is almost three times greater than the speed of the axial rotation of the gainer, which is approximately 85 km/s according to the simulations by \cite{Skulsky1991}. This indicates the certain autonomy of the satellite-disk rotation. Consequently, it should be adopted that the accretion disk consists of two parts around the hot disguised gainer: an external satellite-disk and of the internal thick opaque disk, elongated also in the orbital plane.

It is important to note that the large axis of the satellite-disk deflected in Fig.\, \ref{fig_scheme} outwards from the direction of motion of the gainer by about (20-25)$\degr$. However, it was evident, that in observations data, also taking into account the data of \cite{Struve1941}, \cite{Sahade1966} and other researchers are fixed the long-period changes in the parameters of the satellite-disk. They were confirmed in the following article by \cite{Skulskij1993c} as a result of a comprehensive study of changes in all spectral characteristics of the SiII $\lambda \lambda$ 6347, 6371 doublet, conducted separately for each of the seasons of Beta Lyrae observations 1991-92. According to the result of observations 1992, the rotation of the elongated axis of the satellite disk was detected. For example, Fig.\, \ref{b_lyrae_SiII} shows the development of 
the spectral characteristics of satellite lines in the SiII $\lambda \lambda$ 6347, 6371 doublet from their appearance to disappearance for the observations 1991-92. There is a clear asymmetry in the radial velocity maxima at the 0.958\,P and 0.022\,P phases, which confirms the ellipsoidal shape of the satellite disk. But, the satellite lines in 1992 were clearly observed at the (0.009-0.014)\,P phases, whereas in 1991 they were visible only from the 0.021\,P phase. In 1992, their equivalent widths increased sharply at phases of (0.920-0.925)\,P. Hence, the complex seasonal investigation of the absorption lines of both components and the emission in these SiII red lines showed a clear change in the conditions of the projection of the satellite-disk onto the surface of the bright donor due to the change in their relative configuration \citep{Skulskij1993c}. Over the course of a year, the long axis of the satellite disk has been deflected in the direction of the donor and turned counter-clockwise by about 45$\degr$. This result foresees a rotation of the satellite-disk around the gainer, but it can also reflect precessional motion this disk, in particular, with a period less than one year. In this regard, it should be noted that the article by \cite{Osaki1985} emphasizes that the accretion disk can be deformed coherently into elliptical forms since they are the eigenmodes of such a disk. 

\subsection{Spectral observations of the Beta Lyrae system and masses its star components}
Our spectral observations of the Beta Lyrae system are begun in the middle of the 1960s when already there has been a substantive change of ideas about the evolution of binary systems. From the understanding of the new concept \citep{Crawford1955}, it followed that the hidden secondary component should be more massive, although and underluminous for its mass. Therefore, investigating the complex spectrum of this binary system, apart from the obvious problems of the time, such as studying the chemical composition of the atmosphere of a bright component or elements of its orbit, the task was to find evidence of the line spectrum of the secondary component. From the articles of authoritative researchers of the Beta Lyrae system, it followed that the absorption lines, which would track the motion of the secondary component in the spectrum of this binary system, are absent. However, in the article \cite{Sahade1966} it can be observed that two absorption lines were visible in lines of a red doublet of silicon in addition to satellite lines. This spectral spectral material was concentrated around primary minimum and, probably, therefore, on the question, what can explain the doubling of SiII absorption lines at certain phases, J. Sahade responds as follows: "One of the components gives velocities that agree with the lines of shell which surrounds the whole system, while the other arises from the primary (B8) star". Moreover, the absence of absorption lines of the secondary component in the Beta Lyrae spectrum was confirmed in a detailed review by \cite{Sahade1980}. 

At this time, due to the importance of identifying the spectral lines of the secondary component (then the possibility of a black hole was also considered), a special time was allocated for the Beta Lyrae observations on the 2-m telescope of the Shemakha Astrophysical Observatory. As a result, by \cite{Skulskij1975a} the first determination of the masses of both components was carried out directly from the high-resolution spectrograms (0.1\,$\AA$/mm). Shallow and wide absorption lines of ionized elements, in particular, magnesium, a purple calcium doublet, and a red silicon doublet were detected, the radial velocities of which reflected the orbital motion of the secondary component at a mass ratio of approximately {\it q} = 0.23. The secondary component (currently a gainer) should be at least 4.3 times more massive than the bright donor.

More accurate determination of the orbital parameters of both components, determination of their masses and absolute dimensions of the Beta Lyrae system was carried out in the SiII $\lambda \lambda$ 6347, 6371 lines after a series of CCD observations on the 2.6-m CrAO telescope in 1989-1992 (see \cite{Skulsky1991} with the calculation of synthetic spectra), \cite{Skulskij1992,Skulskij1993c}). The mass ratio of the gainer to the donor reaches a value of 4.47 (with an error less than 0.1) and is about of $q = K_d/K_g = 0.224$ (here $K_d$ and $K_g$ are the semiamplitude of the radial velocities of the donor and the gainer, respectively). At an inclination of the orbit at about $i = 90\degr$ and the mass of the donor in range of (2.9 - 3.0)\,$M_\odot$, the mass of the gainer there is in range of (13 - 13.5)\,$M_\odot$. It is clear that the masses of both components need further consideration and a more precise definition for practical use. 

First of all, one should take into account the somewhat distorted profiles of the red silicon lines of both components due to variable and low ambient emission near these lines. Also, according to \cite{Skulskij1993c}, from the 1991 season to the 1992 season the off-season spectral differences in the lines of SiII $\lambda \lambda$ 6347, 6371 were recorded. In particular, this concerns significant changes in the equivalent widths of the absorption lines both the donor and the gainer. The two-year averages for $K_d = 188.0$\,km/s and for $K_g = 42.1$\,km/s are within the 2\,km/s between 1991 and 1992 with maximum value in the 1992 season. Therefore, at least for the semiamplitude of the radial velocity of the donor, it is reasonable to prefer the less distorted lines of the donor atmosphere in the spectrum of the blue region. Our Beta Lyrae orbital elements obtained from observations on the 1.2-m CrAO telescope over 4 seasons in 1966-76 showed between them some differences, see in \cite{Skulskij1978}. In particular, the average $K_d$ values,  determined from all measured lines of the blue region, are 185.2 km/s, varying within 0.4 km/s; but when determined only due to the 5 SiII lines, they vary from 184.2\,km/s in 1966 and 1969 to 186.8\,km/s in 1972 and 1976, respectively. 

An independent detailed analysis of our radial velocities, together with all published radial velocities over 100 years, was realized, thanks to a thorough investigation of Beta Lyrae orbital elements by \cite{Harmanec1993}. The averaged value for $K_d$ is ($185.9\pm0.33$)\,km/s. There is also averaged value $K_g$ for all measurements of the gainer radial velocities up to 1992 inclusive: ($41.4\pm 1.3$)\,km/s. Then their determination of the component masses and the mass ratio is as follows: $M_g sin^3i = (12.94\pm0.05)\,M_\odot$, $M_d sin^3 i = (2.88\pm0.10)\,M_\odot$ and $q = 0.223$. A similar analysis conducted in \cite{Ak2007} gives for $K_d = (185.27\pm0.20)$\,km/s. The methodologically independent the first astrometric Beta Lyrae observations of \cite{Zhao2008}, taking into account the spectral observations of other researchers, shown the somewhat smaller mass ratio $q = 0.222$ and masses of components of $M_d sin^3i = (2.83\pm018)\,M_\odot$ and $M_g sin^3i = (12.76\pm0.27)\,M_\odot$. In \cite{Mourard2018}, as in the recent modelings of the light curves from the far-ultraviolet to far-infrared regions, was adopted such parameters: $K_d = (186.30\pm0.35)$\,km/s and $K_g = (41.4\pm1.3)$\,km/s at $q = 0.222$. 

Ultimately, the main purpose of such analysis is the need to optimize the masses of both components for further Beta Lyrae studies. Taking into account the previous analysis and all our observational data with some differences during this time, including 1992, the following values of $K_d = 186$\,km/s and $K_g = 41.5$\,km/s may be considered optimal, resulting in $q = 0.223$ and $M_g / M_d = 4.48$. Given the binary orbital inclination of $i = 92\degr$ and $i = 93.5\degr$, which were obtained respectively in \cite{Zhao2008} and \cite{Mourard2018}, and, given all of the above, it is now possible to consider as justified for to the use in the scientific work of the following values: 2.9\,$M_\odot$ for the donor and 13\,$M_\odot$ for the gainer. These masses of components illustrate Fig.\, \ref{fig_scheme}. Considering this result as established, we used the donor mass of 2.9\,$M_\odot$ in \cite{Skulskyy2019} to calculate the inner structure of the donor, which precedes the formation of the degenerate dwarf. It was shown that the mass such degenerate core of the donor should be in the range (0.3-0.5)\,$M_\odot$. 

In view of the foregoing, it is clear that the measurement of radial velocities had some difficulties due to the specificity of the physical conditions in the deformed donor atmosphere. The important role of the chemical composition of the donor atmosphere at the presence such active mass losses should also be emphasized \citep{Skulskij1975b}. The donor absorption lines in the visible region are belonged mainly of the B8III type star, but correspond to a broader excitation and ionization range and vary substantially with the orbital phase. The simultaneous presence of high excitation lines NII, CII, SiIII, FeIII and a group of low excitation lines FeI, FeII, TiII, CrI, CrII, and others are observed. A study of the dependence of chemical abundance on excitation conditions within the limits, which could exist in the donor atmosphere, has led to the conclusion that excess He, N, Si were obtained with all reasonable assumptions about the excitation conditions \citep{Skulskii1986}. The excess of He under the deficiency of H  and the excess of N in comparison with C and O indicate that the atmosphere of the donor substance was significantly reworked during the combustion of hydrogen through the CNO cycle in the bowels of the donor (this is consistent with the findings of \cite{Balachandran1986}). In addition, according to \cite{Skulskij1975b, Skulskii1986} variations in the micro-turbulent velocities from 5.5\,km/s in quadratures to 18\,km/s on the side of the donor facing the gainer indicates the change in the atmospheric density along a deformed quasi-ellipsoidal surface of the donor at a practically filling of its Roshe cavity. In this regard, it is interesting to note that the donor magnetic field was detected with its maximum on the side of the donor facing the gainer \cite{Skulskij1985a}. In particular, the anomalous excess of certain chemical elements on the donor surface is consistent with the magnetic Ap stars.

\subsection{Spectral investigations and the magnetic field of the Beta Lyrae system}
Magnetism as a phenomenon is widely studied in modern astrophysics. The basic question of how the magnetic field of the stars arose and evolved remains unclear. It is more plausible that an anomalous chemical composition on the surface of the stars is created during the evolution of the stars \citep{Romanyuk2005}. The fraction of binary magnetic stars among CP-stars is insignificant, and the detailed study of their physical parameters and the identification of generation mechanisms and patterns in the behavior of their magnetic fields are limited by several objects \citep{Skulsky2009}. From the point of view of stellar magnetism, the detection of a magnetic field in the Beta Lyrae system had of particular interest. It is the only interacting binary system, in which a magnetic rotator protrudes a giant of B-type that has reached its Roche cavity and is actively losing matter. The importance of this phenomenon requires the analysis of all known observations and investigations of the magnetic field.

\begin{figure}[!h]
	\centerline{\includegraphics[width=0.75\textwidth,clip=]{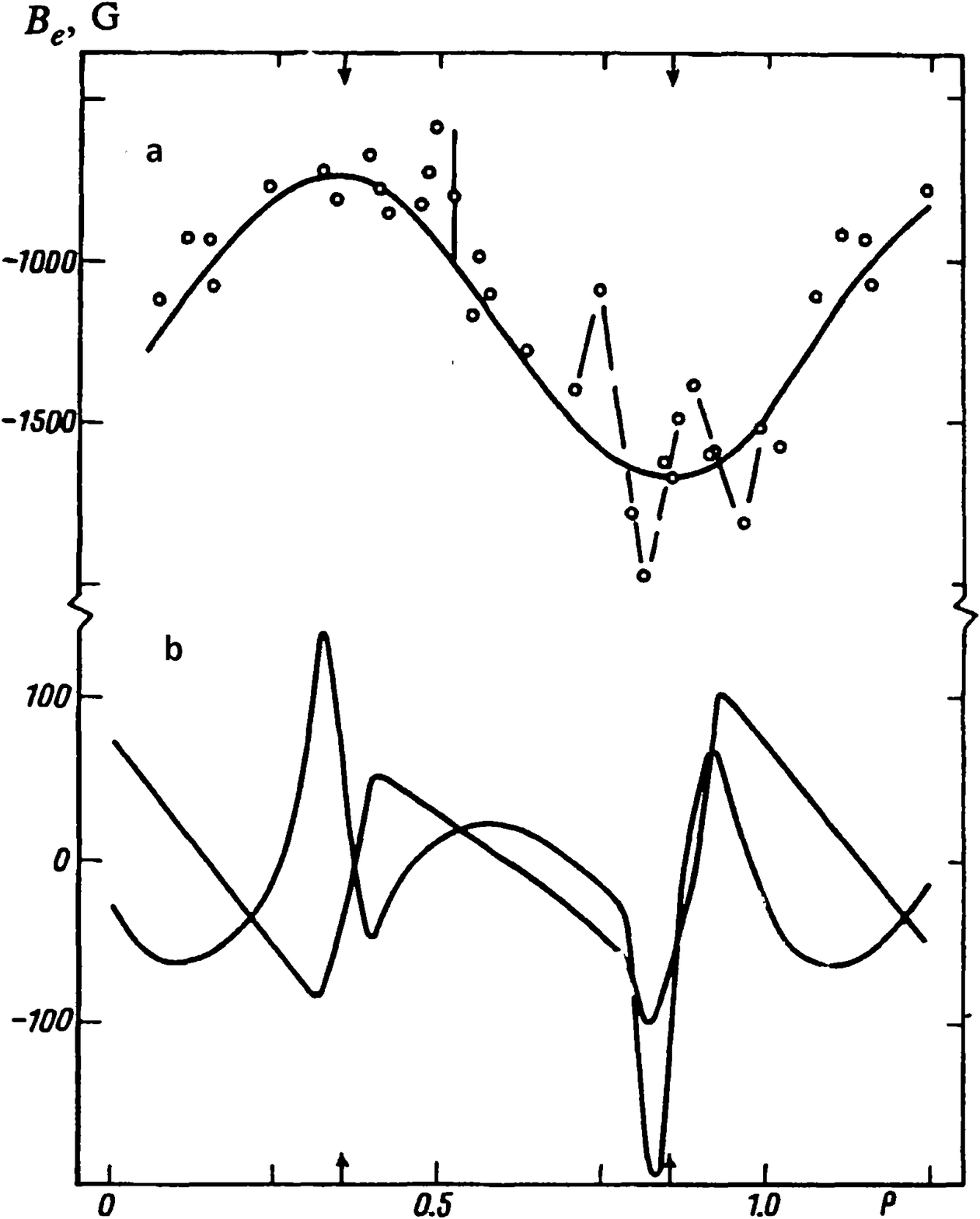}}
	\caption{The photographic curve of effective magnetic-field strength plotted using data of \cite{Skulskij1990}. A characteristic error of measurements shown by vertical bar (2\,$\sigma$). Sinusoidal mean curve attests to the dipole field with poles at phases of 0.355 and 0.855\,P (their centers are shown by arrows). In the region of visibility of the second pole, facing the massive gainer, rapid variations of the field are suspected (a). The averaged assumed $B_e$=$f(P)$ curves for both silicon lines from Figure \ref{fig:6} are given, in order to compare them with one another and with the photographic $B_e$-curve (see text) (b).}
	\label{fig:5}
\end{figure}

The first consistent study of the Beta Lyrae magnetic field was carried out by \cite{Skulskij1982, Skulskij1985a, Skulskij1990} using four hundred of Zeeman spectrograms with 9\,$\AA$/mm dispersion in the blue spectral region that were obtained over 31 nights in 1980-88 on the 6-m telescope of the Special Astrophysical Observatory (SAO). Zeeman splitting was measured in the atmospheric lines of a bright donor. These photographic observations have shown that the effective magnetic field strength varies quasi-sinusoidally over the orbital period, with the amplitude $A=(475\pm51)$\,G and the negative average value of $H_e=(-1198\pm51)$\,G. In order to elucidate the configuration of the magnetic field, in \cite{Skulskij1985a} the first model calculation was performed. The agreement between the theoretical and the averaged observational phase curves for $H_e$=$f(P)$ was achieved as follows: a) the magnetic dipole axis inclined to the orbit plane of the binary system by an angle $28\degr$; b) the center of the magnetic dipole displaced by 0.08\,A from the donor center toward the gainer center (here A is the distance between centers of gravity of both components, see Fig.\,\ref{fig_scheme}). In \cite{Burnashev1991B}, based on the observations of all 31 nights in 1980-88, the phase angle between the line joining the centers of the components and the projection of the magnetic dipole axis on the orbital plane is determined near $52\degr$, and this axis is directed along with the orbital phases of (0.355-0.855)\,P. It would be noted that, at observing the Beta Lyrae system at near these two phases, magnetic poles at the donor surface can be reflected in the magnetic field curve. The maximum effective magnetic field strength is concentrated at the phases of the second quadrature and, importantly, on phases near 0.855\,P of just that one magnetic pole, which is on the donor surface facing the gainer (see Fig.\,\ref{fig:5}).

The spatial structure of the donor magnetic field in its change with the orbital period aroused some interest, but, first of all, all this required independent confirmation. There is an independent argument of the reality of the magnetic field, which is characteristic of the magnetic stars: it is the depression at the $\lambda$ 5200\,$\AA$ in their continuums. The phenomenon of depression at the $\lambda$ 5200\,$\AA$ in the continuous spectrum of the Beta Lyrae system was detected and studied due to the absolute spectrophotometric data for 1974 and 1984 by \cite{Burnashev1986}. They showed that the variability of the equivalent width of depression at 5200\,$\AA$ clearly correlates during the orbital period with the variability of the magnetic field: both maxima of the equivalent width of this depression are observed at the same phases as the zones of both poles of the magnetic field on the donor surface. 

\begin{figure}[!h]
	\centerline{\includegraphics[width=0.75\textwidth,clip=]{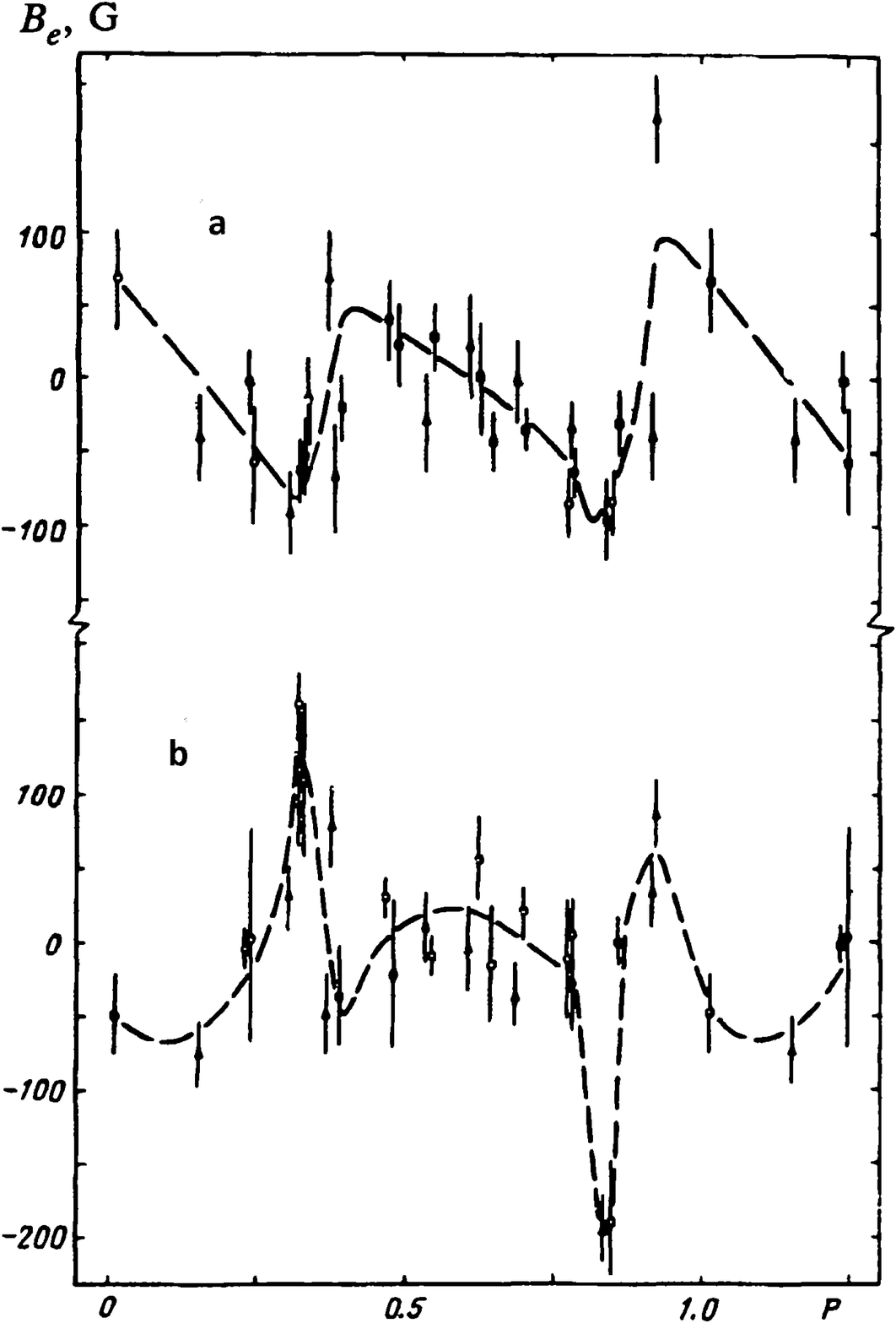}}
	\caption{
	The phase variability of effective intensity of the longitudinal component of the magnetic field in SiII absorption lines $\lambda$  6347 (a) and $\lambda$ 6171 (b) of bright component, based on a study of Zeeman CCD spectrograms. Vertical error bars correspond to 2\,$\sigma$. Dashed curves were drawn by hand on the basis of assumed $B_e$=$f(P)$ variability.}
	\label{fig:6}
\end{figure}

The next were measurements of the Zeeman splitting at SiII $\lambda \lambda$ 6347, 6371 lines by \cite{Skulsky1993}. They were carried out in 1991-92 with a Stokesmeter and CCD detector mounted on the 2.6-m CrAO telescope. These fairly homogeneous observations in all orbital phases are presented in Fig.\,\ref{fig:5} and \ref{fig:6} (as copies of Figures 1 and 2 from the article of \cite{Skulsky1993}). 

The effective intensity of the longitudinal component of the magnetic field varies over an orbital period within $\pm$200\,G, however, the shape of the curves and their amplitudes for each of the silicon lines are slightly different, although they somehow reflect the photographic curve of the magnetic field (as of field of a mainly dipolar configuration). Fig.\, \ref{fig:6} shows, in particular, that changes in the magnetic field are clear only in the phases near the poles of the magnetic field, whereas in the phases near 0.5\,P (donor in front of the observer) and in phases near 1.0\,P (the donor is closed by the accretion disk), the magnetic field is practically near zero. Moreover, the average magnetic field for 499\,days of these CCD-observations is near -11G. However, it can be seen from Fig.\, \ref{fig:5} and \ref{fig:6} that the polarity of the magnetic field clearly changes in the phases (0.32-0.39)\,P of the first quadrature and in the phases (0.82-0.89)\,P of the second quadrature. In both lines, these changes occur within 0.1\,P, but in the central regions of the magnetic field poles, that is, exactly in the phases 0.355\,P and 0.855\,P, the field passes through zero. The variability of the magnetic field occurs slightly differently at both poles, however, Fig.\,\ref{fig:5} and \ref{fig:6} show that the behavior of both magnetic field curves near the 0.855\,P phases is similar (i.e., in the phases of observing the magnetic pole facing the gainer). Moreover, the behavior of the photographic curve of the magnetic field is also similar in the limits of these (0.81-0.88)\,P phases. This means that the physical conditions of the formation of the magnetic field along the transverse dimensions of this magnetic pole pole can differ substantially from two sides from the center of this pole.

Hence, the measurements of the Zeeman splitting at SiII $\lambda \lambda$ 6347, 6371 lines in \cite{Skulsky1993} have shown that a magnetic field is formed on the donor surface mainly around the poles of the magnetic field, locally changing the polarity of the magnetic field. These CCD observations allow the identification of both poles of the magnetic field on the donor surface. The transverse dimensions of both poles are given by the phase difference between the phases of change in the polarity of the magnetic field in both lines. The average of the phase difference is about 0.08\,P. These observations independently confirm the presence of a hot region on the donor surface detected by the infrared photometry of \cite{Zeilik1982} in the phases (0.8-0.9)\,P (i.e., at the observation of the magnetic pole facing the gainer). This hot region was also detected in \cite{Skulskij1993c} due to the deep eclipse of the region of the SiII emissions formation on the donor surface by the satellite disk surrounding the gainer. It should be recalled that the SiII $\lambda \lambda$ 6347, 6371 absorption lines are limited by weak emission. The energy and dynamic characteristics of the absorption and the emission components of these lines show their variability, which is correlated synchronously with such variability of the donor magnetic field \citep{Skulskij1993a, Skulskij1993c}. This study also has shown that both components of the SiII $\lambda \lambda$ 6347, 6371 lines are formed in a close spatial environment near the donor atmosphere, which can reach to the Roche cavity. \cite{Skulsky1993} also noted that the magnetic field becomes less intense with increasing the height in the donor atmosphere.

In 1993-1995 and 2000-2004, CCD observations and Zeeman splitting measurements were continued,but in some metallic lines in the blue spectral region \citep{Skulsky2005}. Although these observations could not fully cover the orbital period, in general, the magnetic field demonstrated the time-dependent complicated behavior. For example, the arithmetic means value for a magnetic field for 20 observation dates 1993-1995 is $(285.3\pm95.5)$\,G. In 1993-2004 five observation dates were obtained showing significant values of the magnetic field with the ratio $k = H_e /\sigma>$(3-6) in the $H_e$ ranges from -812\,G to 1348\,G. Special testing has shown that such registered magnetic field values are reliable. The possibility of the orbital and longer time-scale variability of the Beta Lyrae magnetic field is proposed in \cite{Leone2003} based on the dynamo mechanism. It should be noted that independent observations of the effective magnetic field, carried out over a 13-day full orbital period in June 1999 at the Observatorio Astrofisico di Catania (OACt), 
revealed a magnetic field of positive polarity with an average value of $(1.288\pm163)$\,kG. However, within the accuracy range, these measurements do not show any obvious changes in the magnetic field during the orbital period, and such OACt observations were no longer repeated. 

It should be noted that \cite{Skulsky1993} also obtained the first data, which may indicate the
presence of a magnetic field in the pseudophotosphere of a thick accretion disk surrounding the
gainer. It cannot be excluded that the magnetic field of the donor is in some way magnetically
related to the magnetic field generated by the accretion disk, or that changes in the donor
magnetic field may correspond to variations in the structures of the accretion disk. The
frequency analysis of the magnetic field measurements did not reveal obvious periodic processes
with long-term-scale variability, for example, with the well-known 283-day period. However, an
agreed system of secondary periods and resonances had been identified, which links the
substructures of the accretion disk and the surface magnetized structures of the donor
\citep{Skulsky2007}. 
The presence of a common magnetosphere, which can manifest itself in the circumstellar gas structures, cannot be excluded. Moreover, according to \cite{Shore1990} in helium stars with a magnetically controlled matter (the atmosphere of the donor in the Beta Lyrae system also has excess helium), magnetized plasma can be directed outward from the magnetic polar regions on the star surface, forming the magnetosphere. 

As follows from the foregoing analysis, all existing observations and measurements of the magnetic field indicating its presence in the donor atmosphere do not form the single magnetic field curve variability. 
It cannot rule out of a long time magnetic field variability superimposed on the orbital period. Only two curves, the $H_e$=$f(P)$ as photography curve in the blue region of spectrum (obtained from 1980-1988 observations on the 6-m SAO telescope) and such curve in the red region of spectrum (obtained from 1991-1992 CCD observations on the 2.6-m CrAO telescope in the silicon lines), show a certain correlation in the changes of the magnetic field over the orbital period. These CCD observations have shown that the magnetic field becomes apparent only at phases around the poles of the magnetic field obtained from the solution of the photographic curve of the magnetic field. And the observations on the 6-m SAO telescope lead to an average value of the longitudinal magnetic field, which is significantly overestimated relative to such values in CCD observations. However, the quasi-sinusoidal photographic curve of the magnetic field looks as one would expect with respect to magnetic stars. Thus, although the average magnetic field measured on the 6-m SAO telescope may be overestimated, it should be considered such variability of the donor magnetic field important for further studying the mass loss and accretion pattern in the Beta Lyrae system.

\section{Short conclusion and outlook}
The phase variability of the donor magnetic field and its specific configuration led to evident research of their interconnection with the configuration of the accretion gas structures in the Beta Lyrae system. This task remained relevant during long term observations and studies of the magnetic field. In the course of further processing of a large amount of research material, it became clear that there were a number of different observable facts that needed systemic coverage. This required a generalization of the current state of diverse, primarily, spectral studies of the Beta Lyrae system. This idea worked.  In the research conducted in this article, several important directions of work should be noted as of a definite basis for studying the processes of mass transfer in the presence of the magnetic field. 

In order to optimize the mass of both components for use in further scientific work, spectral and astrometric observations and model calculations of light curves were also analyzed. 
As a result, it can be argued that the currently optimal masses for the Beta Lyrae system are the following values: 2.9\,$M_\odot$ for the donor and 13\,$M_\odot$ for the gainer (they used to illustrate Fig.\,\ref{fig_scheme}). 

The study of satellite lines as a phenomenon in the Beta Lyrae spectrum led to the conclusion that the accretion disk consists of two parts: an outer satellite disk and an internal massive opaque disk, which have an ellipsoidal shape. The stratification of the excitation conditions of the various satellite lines shows an increase in temperature and an acceleration of the rotation of the outer layers of the disk to the center of the gainer. A possible precessional movement of the accretion disk leads to its coherent deformation into elliptical forms as the eigenmodes of this disk. It turned out that the study of the spatial structure of the accretion disk and the secondary non-orbital periods of this binary system (e. g., the well-known of the 283-day period) should be carried out jointly taking into account the spatial configuration of the donor magnetic field. This led to forming a coherent system of secondary periods and resonances that change synchronously with the growth of the orbital period. 

From the analysis of all observations and studies of the magnetic field, the long time variability of the magnetic field, in general, is still not clear. This phenomenon should be investigated more fully. It is turned out, that the systematic observations of 1980-1988 on the 6-m SAO telescope could be considered of the most reliable data. They formed a donor magnetic field curve over a large number of orbital periods, which looks like such curves for magnetic stars. The spatial configuration of the donor magnetic field performed owing to the model calculation, is suitable for further research of the question on the correlation between the orbital variability of the magnetic field and such variability of physical parameters in different processes, which take place between the components of the binary system. It is known, that the structure of gas flows in the processes of accretion and mass exchange between the components of the binary system changes in a certain way over the orbital period. If these processes are reflected in the orbital variability of the magnetic field or vice versa, then it is necessary to consider their mutual influence and investigate as a certain phenomenon.       

The picture of the accretion of magnetized flows as a result of the preliminary study of the above correlation processes can be understood from \cite{Skulsky2015, Skulsky2018}. It has been found that the spatial configuration of the donor magnetic field influences the formation of magnetized gas structures between the components of the binary system. The collisions of the magnetized plasma with the accretion disk, surrounding the gainer, are especially effective in phases of the second quadrature when the heating of the accretion disk is greatly enhanced. This heating is also enhanced by the rapid counter-rotation of the accretion disk. The spatial configuration of the magnetic field of the donor is reflected by different physical processes in the accretion structures of this binary system. Further studies show this picture more fully, but this requires shedding light on a number of different observable facts from the far ultraviolet to the red region. Relevant materials are being prepared for coverage soon. 

\acknowledgements
The author is thankful to Plachinda S.I for constructive comments and Kudak V.I. for consultations.

\bibliography{Skulsky}
\end{document}